\documentclass[fleqn,usenatbib,useAMS]{mnras}

\usepackage{newtxtext,newtxmath}
\usepackage[T1]{fontenc}
\usepackage{ae,aecompl}

\DeclareRobustCommand{\VAN}[3]{#2}
\let\VANthebibliography\thebibliography
\def\thebibliography{\DeclareRobustCommand{\VAN}[3]{##3}\VANthebibliography}

\usepackage{graphicx} %For figure placement
\usepackage{amsmath} %For mathematics typesetting
\usepackage{physics} %Provides additional typesetting options and shortcuts for common mathematical expressions
\usepackage{color,soul,framed,xcolor} %For marking revisions

\newcommand{\fwhm}{\text{FWHM}} %Simple short cut for writing FWHM within a mathematical expression
\defcitealias{Cohen_X-ray_mass_loss_method}{C2010}
\defcitealias{Cohen_zeta_pup_mass_loss_2020}{C2020}

\title[Hot Star Mass-loss Rate Testing]{Testing the Reliability of X-rays as a Tool for Constraining Mass-loss Rates of Hot Stars}

\author[S. J. Gunderson et al.]{Sean J. Gunderson$^1$\thanks{Contact e-mail: \href{mailto:sean-gunderson@uiowa.edu}{sean-gunderson@uiowa.edu}},
Kenneth G. Gayley$^1$,
Pragati Pradhan$^2$,
David P. Huenemoerder$^2$,
Nathan A. Miller$^3$
\\
$^1$Department of Physics and Astronomy, University of Iowa, Iowa City, IA 52242, USA\\
$^2$Massachusetts Institute of Technology, Kavli Institute for Astrophysics and Space Research, 77 Massachusetts Ave., Cambridge, MA 02139, USA\\
$^3$Department of Physics and Astronomy, University of Wisconsin–Eau Claire, Eau Claire, WI 54701, USA\\
}

\date{Accepted 2022 February 22. Received 2022 February 16; in original form 2021 September 20}

\pubyear{2022}

\begin{document}
\colorlet{shadecolor}{yellow}
\label{firstpage}
\pagerange{\pageref{firstpage}--\pageref{lastpage}}
\maketitle

%%%%%%%%%%%%%%%%%%%%%%%%%%%%%%%%%%%%%%%%%%%%%%%%%%%%%%%%%%%%%%%%%%%%%%%%%%%%%%%%%%%%%%%%%%%%%%%%%%%%%%%%
%%%%%%%%%%%%%%%%%%%%%%%%%%%%%%%%%%%%%%%%%%%%%%%%%%%%%%%%%%%%%%%%%%%%%%%%%%%%%%%%%%%%%%%%%%%%%%%%%%%%%%%%

\begin{abstract}
    We fit a new line shape model to \textit{Chandra} X-ray spectra of the O supergiant $\zeta$ Puppis to test the robustness of mass-loss rates derived from X-ray wind line profiles against different assumed heating models. Our goal is to track the hot gas by replacing the common assumption that it is proportional to the cool gas emission measure. Instead of assuming a turn-on radius for the hot gas (as appropriate for the line-deshadowing instability internal to the wind), we parametrize the hot gas in terms of a mean-free path for accelerated low-density gas to encounter slower high-density material. This alternative model is equally successful as previous approaches at fitting X-ray spectral lines in the 5 -- 17 \AA\ wavelength range. We find that the characteristic radii where the hottest gas appears is inversely proportional to line formation temperature, suggesting that stronger shocks appear generally closer to the surface. This picture is more consistent with pockets of low-density, rapid acceleration at the lower boundary than with an internally generated wind instability. We also infer an overall wind mass-loss rate from the profile shapes with a technique used previously in the literature. In doing so, we find evidence that the mass-loss rate derived from X-ray wind line profiles is not robust with respect to changes in the specific heating picture used. 
\end{abstract}

%%%%%%%%%%%%%%%%%%%%%%%%%%%%%%%%%%%%%%%%%%%%%%%%%%%%%%%%%%%%%%%%%%%%%%%%%%%%%%%%%%%%%%%%%%%%%%%%%%%%%%%%
%%%%%%%%%%%%%%%%%%%%%%%%%%%%%%%%%%%%%%%%%%%%%%%%%%%%%%%%%%%%%%%%%%%%%%%%%%%%%%%%%%%%%%%%%%%%%%%%%%%%%%%%

\section{Introduction}

X-ray spectra from hot stars (spectral types O, B, and WR) are a well established diagnostic for constraining stellar wind properties. A significant example of this is the technique developed in \citet[hereafter \citetalias{Cohen_X-ray_mass_loss_method}]{Cohen_X-ray_mass_loss_method} where the authors determined the mass-loss rate of the O supergiant $\zeta$ Puppis (HD 66811) from the shape of its X-ray line profiles. The line shapes constrain the total wind column, and hence mass-loss rate, while also providing information about the spatial distribution of the heating. However, the precise nature of the spatial distribution can be parametrized in very different ways, depending on the assumed efficiency of the heating mechanism. The approach of \citetalias{Cohen_X-ray_mass_loss_method} was inspired by the line-deshadowing instability (LDI, e.g., \citet{Owocki_LDI}; \citet{Owocki_LDI_Porosity}), which is less active in the hemispheric radiation field near the surface. Hence they fit to a turn-on radius and were able to achieve satisfactory agreement with the hard X-ray profile shapes. We will refer to this general approach to hot-gas parametrization as the LDI-inspired version of an Embedded Wind Shock Model (LDI-EWSM).

When dealing with low-order parametrizations, however, a successful fit does not imply uniqueness of the model, and other pictures for how the heating occurs might also motivate parametrizations that can fit the X-ray line shapes. This paper considers such an alternative, that replaces a turn-on radius with a concept of a velocity-dependent mean-free path for fast gas to overtake the slower prevailing wind. Also, the \citetalias{Cohen_X-ray_mass_loss_method} approach is to assume the hot-gas emission measure is proportional to that of the cool unshocked wind, consistent with a picture of hot gas appearing over a radially constant fraction of wind volume. But there is no a priori reason for the hot gas to exhibit such a radially constant volume filling factor, especially when the filling factor is locally regulated by radiative cooling of the shocked gas, so the approach taken here is to assume efficient radiative cooling that makes no necessary reference to the emission measure of the cool gas. When all the heat thermalized in the shocks is assumed to radiate quickly, the hot-gas emission measure simply responds to the local heating rate, and it is only the latter that requires parametrization. This contrasts with the \citetalias{Cohen_X-ray_mass_loss_method} approach, which is more natural for inefficient radiative cooling that pegs the X-ray flux to the predetermined emission measure of the cool gas. 

In summary, our goal is to perform an analysis similar to \citetalias{Cohen_X-ray_mass_loss_method} on recently updated \textit{Chandra} X-ray data, but for a heating parametrization that is more appropriate for fast flows that are initiated from the lower boundary and which create shocks that efficiently radiatively cool, whereas the \citetalias{Cohen_X-ray_mass_loss_method} approach uses an X-ray generation that is more appropriate for inefficiently cooled shocks created internally in the wind by the LDI beyond an assumed initiation radius. Our goal is to test whether such an alternate parametrization can also successfully fit the profiles, and if doing so yields additional insights about the heating mechanism. We also wish to test how sensitive the inferred wind column depths are to the choice of heating parameterization. Our confidence in the inferred mass-loss rates will be improved if the results are not strongly contingent on the choice of parameterization.

The \citetalias{Cohen_X-ray_mass_loss_method} technique uses the \textit{windprof} and \textit{hewind} local models\footnote{\url{https://heasarc.gsfc.nasa.gov/xanadu/xspec/models/windprof.html}} in \textsc{xspec} to fit X-ray line data. These two models provide a fiducial optical depth value for each line that is used to fit a wavelength-dependent opacity curve with the mass-loss rate as a free parameter. Since this mass-loss diagnostic is based on X-ray absorption over the global wind column depth, it smooths over local density fluctuations and is thus insensitive to the uncertainties of wind clumping. The efficacy of the technique was verified in \citet{Cohen_mass_loss_archive}, where, for an ensemble of O stars in the \textit{Chandra} archive, the authors found mass-loss rates consistent with H$\alpha$ measurements with an assumed clumping factor of $f_{cl} = 20$. Of particular note is the mass-loss rate of $\zeta$ Pup, which was found to be $\dot{M}=1.8\times10^{-6}$ M$_{\sun}$  yr$^{-1}$; a confirmation of multiwavelength observations. 

The data set used for this original mass-loss-rate determination was taken by \textit{Chandra} during Proposal Cycle 1. In addition to the question of how robust is the result to different heating assumptions, there is also the question of how stable are the mass-loss determinations over observing epochs, and whether mass-loss rates can actually change in real time.  This latter question was addressed by the authors of \citet[hereafter \citetalias{Cohen_zeta_pup_mass_loss_2020}]{Cohen_zeta_pup_mass_loss_2020}, using recent 813 ks \textit{Chandra} observation data from Proposal Cycle 19.  Surprisingly, they found a new mass-loss rate of $\dot{M}=2.47\times10^{-6}$ M$_{\sun}$  yr$^{-1}$, corresponding to a 40\% increase in just under two decades. As noted by the authors of \mbox{\citetalias{Cohen_zeta_pup_mass_loss_2020}}, changes to $\zeta$ Pup's mass-loss rate are expected to be only a few per cent, due to observed low-level stochastic and periodic variability. Thus the much larger inferred change suggests either an unnoticed gradual increase or a sudden jump due to some as-yet unidentified change in $\zeta$ Pup's stellar properties. Unfortunately, the \textit{Chandra} observations of $\zeta$ Pup do not cover the period over which a steady change could have occurred, so the nature of the change, if real, remains mysterious.  This further underscores the importance of testing the robustness of this technique for inferring mass-loss rates, including the use of a different heating parametrization, as done here.

Notwithstanding the issues involving mass-loss rate, the main purpose here is to use the X-ray line shapes to constrain the spatial distribution of the heating. In particular, we wish to find clues that can distinguish whether the heating comes from rapid flows that accelerate quickly from the surface, as if from gas streams that are fundamentally underloaded from the start, or if it exhibits a ``turn-on'' distance that is indicative of inefficient LDI action close to the surface. Hence the key question is whether the LDI can explain the spatial distribution of X-rays that \citetalias{Cohen_X-ray_mass_loss_method} -- \citetalias{Cohen_zeta_pup_mass_loss_2020} have found, as  discussed further in \S \ref{sec:discuss}. 
 
In \citet{Feldmeier_LDI_Non-linear_Growth}, it was shown that the LDI can take a tiny photospheric perturbation (e.g., a short-wavelength low-amplitude acoustic wave) and stretch it to globally large scales that lead to nonlinear shock structures. Importantly, in the authors' scale-free model (which for simplicity does not include gravity but still provides important lessons in the functioning of the LDI), they find that velocity disparities grow linearly with height $\Delta v \sim z \sim v$. Since velocity tracks with the radius, we would then expect that the strength of the shocks would rise with distance from the surface, suggesting a positive correlation between turn-on radius and X-ray temperature. Recent MHD LDI simulations by \citet{Driessen_LDI_Sim} have also shown this to be the case. For the case of zero magnetic field, their simulated shock structures grow in strength until saturating at a peak between 2 -- 3 stellar radii. Thus we would expect the hardest lines produced by the LDI to appear at similarly large radii, while soft lines could exhibit much smaller turn-on radii. We wish to test this with the X-ray profile data by contrasting with an essentially opposite view, where stronger shocks stem from rapid acceleration of streams that are underloaded right from the stellar surface and appear at lower radius because they overtake the prevailing wind more rapidly.

This rapid overtaking is an alternative class of acceleration mechanisms we refer to as Variable Boundary Conditions (VBC). For these mechanisms, the velocity disparities that lead to the X-ray emitting shocks are due to postulated local and transient changes in the initial wind acceleration at the photosphere, providing an alternate explanation for embedded wind shocks (VBC-EWSM). An example of such a VBC would be if the mass loading of the wind is variable across the stellar surface; either spatially or temporally. If such a variability were to exist, under-loaded portions would have a much steeper acceleration profile due to reduced self-shadowing of the line driving, a nonlinear effect that is distinguishable from the LDI.  The analysis of \citet{Gayley95} concludes that such reduced self-shadowing could, in an extreme case, increase the line acceleration by up to three orders of magnitude.  This could shorten the acceleration length scale by the square root of that, or several factors of 10, thus able to achieve speeds at the scale of 1000 km s$^{-1}$ in a small fraction of a stellar radius before encountering a shock.  Furthermore, the strongest shocks would be from the most underloaded and fastest accelerated material, which would as a consequence appear at the smallest radii, and this is what we wish to test.

One may question such surface variabilities on the grounds that the wind solution should be stable like that of the solar wind. For example (as argued by Achim Feldmeier in private correspondence), proposed variabilities could be corrected by inwards travelling waves (e.g., Abbott waves \citep{Feldmeier_Abbott_waves} that would return the subcritical under/over-loaded flows to the stable solution. However, this assumes that wind initiation is mathematically similar to that of the Sun, analyzable as an X-type critical point.  Detailed line-driven wind simulations by \citet{Sundqvist_topology} have shown that including the effects of scattering, something that is of non-negligible importance in the forcing on the ions, causes the topology of the critical point to be of nodal type, which produces a family of degenerate solutions that the system may take. For a realistic case of an Eddington-limb-darkened source, \citet{Sundqvist_topology} find that this nodal type topology causes strong variabilities in the wind that extend down to the photosphere. Moreover, any variability at the inner boundary (e.g., bright spots \citep{Ram_Brite_Spots_ZP}, g-modes, or radiation-driven magneto-acoustic waves \citep{Fernandez_RMI, Sen_RMI_in_stars}) would add to this variability and potentially cause shock structures to appear quickly after the wind leaves the surface.

Nevertheless, the LDI is expected to operate, so we are not attempting to remove or disprove the LDI; instead we are exploring the possibility that the strongest shocks (and hence hardest X-rays) may come from a different mechanism, important in the lower wind. Hence this work should be viewed as exploring how a VBC model would affect X-ray line profile shapes, using a low-order heating parametrization to study the connection of those parameters with the overall profile characteristics of line width and line shift.

Notwithstanding the aforementioned differences in heating parametrization, our approach to fitting the line profiles follows closely that of \citetalias{Cohen_X-ray_mass_loss_method} -- \citetalias{Cohen_zeta_pup_mass_loss_2020}, as they define a spectral line model fit using a triplet of parameters ($\tau_*, R_0, N$). The last parameter, $N$, is merely a normalization that relates to how many shocks are present, so is not of importance to our analysis because we are not studying the line flux magnitudes (and indeed a study of the line flux ratios has already been undertaken by \citet{Dave_zeta_pup_data}).  Our focus is on the profile shapes, which relate only to the other two parameters. The $\tau_*$ parameter is referred to as a fiducial optical depth, i.e., a characteristic optical thickness that a photon would have traveled through on a central ray that originates at the bottom of the wind when $v = v_\infty$. The parameter $R_0$ is the radius at which X-ray production begins, i.e., the previously mentioned ``turn-on" radius, in units of the stellar radius. See \citet{Cohen_model_behavior} for how these parameters affect their model. Our own analysis differs only in how we treat the spatial distribution of the hot gas, so we replace the $R_0$ parameter\footnote{In the documentation for \textit{windprof} and \textit{hewind}, $R_0$ is given in the inverse radius parameter $u_0$.} with a related length scale $\ell_0$ defined below. In both our approach, and the Cohen approach, all these parameters are line dependent.

This paper is organized as follows. In \S \ref{sec:model}, we derive and explain the different pieces of our model while comparing and contrasting relevant aspects with the reference model. In \S \ref{sec:modelbehavior}, we give example plots of our model and detail the behavior of relevant observables as parameters are varied. In \S \ref{sec:data}, we provide details on the data, its reduction, and the fitting procedure. Finally, in \S \ref{sec:results} and \S \ref{sec:discuss} we discuss our results and their implications.
%%%%%%%%%%%%%%%%%%%%%%%%%%%%%%%%%%%%%%%%%%%%%%%%%%%%%%%%%%%%%%%%%%%%%%%%%%%%%%%%%%%%%%%%%%%%%%%%%%%%%%%%
%%%%%%%%%%%%%%%%%%%%%%%%%%%%%%%%%%%%%%%%%%%%%%%%%%%%%%%%%%%%%%%%%%%%%%%%%%%%%%%%%%%%%%%%%%%%%%%%%%%%%%%%

\section{Our Model}\label{sec:model}

We begin with the assumption that the stellar wind around a hot star is a 2-component plasma made of parcels of slow gas, commonly referred to as ``clumps" in the literature, and faster ``streams" that collide with the slower clumps. To model an X-ray spectral line, we define the normalized line-of-sight Doppler shift as $\xi \equiv - \mu v(r) / v_\infty$, where $v(r)$ is the local wind speed at the point where the X-ray emission occurs, $\mu$ is the direction cosine to the observer, and $v_\infty$ is the terminal speed of the wind. Hence $\xi$ determines the contribution of some region of the wind to the unitless location in the X-ray profile as a fraction of maximum Doppler shift, where $\xi < 0$ implies moving toward the observer (i.e., the forward hemisphere) and $\xi = 0$ corresponds to the unshifted rest wavelength of the line.

The average, unshocked clumps of the wind are assumed to follow the usual $\beta$-velocity law
\begin{equation}
    v(r) = v_\infty\left(1-\frac{R_*}{r}\right)^\beta
\end{equation}
with $\beta=1$ is chosen because it allows for analytical solutions to equations such as the one for optical depth
\begin{equation}
    \tau(\mu,r;\tau_*)=\tau_*(\lambda)\int_0^\infty \frac{R_*v_\infty}{r'^2v(r')}\dd s,\label{eq:tauformula}
\end{equation}
where the integration variable $s$ is the path of the photon's travel. The variable $\tau_*(\lambda)=\kappa(\lambda)\dot{M}/4\pi R_*v_\infty$ is the first parameter of our model and defined identically as the parameter of the same name as models used in \citetalias{Cohen_X-ray_mass_loss_method} -- \citetalias{Cohen_zeta_pup_mass_loss_2020}.

To generate a model line profile, we exploit the assumed spherical symmetry in that all observers in all directions should see the same profile. This allows us to sum, over the entire column depth of the wind, the emissions from any streams that overtake slower gas with sufficient speed to produce the X-ray line in question, keeping track of the appropriate Doppler shift $\xi(\mu)$ for the wind speed at the location of the emission. There is thus no reason to specify the gas density at the location of the emission, all we need to parametrize is the location and post-shock speed where the relevant shock occurs. The differential emission within a Doppler bin for our model is thus given by
\begin{equation}
    L(\xi;\tau_*,\ell_0)=\int_{r_{m}(\xi)}^\infty \dv{\mu}{\xi}\mathrm{e}^{-(r-R_*)/\ell_0}\mathrm{e}^{-\tau(\mu,r;\tau_*)}\dv{P}{r}\dd r, \label{eq:luminositybase}
\end{equation}
where
\begin{equation}
    \dv{\mu}{\xi} = \frac{1}{2}\frac{v_\infty}{v(r)}.
\end{equation}
We detail each piece of this model below.

For the optical depth, we first note the orientation we have chosen. The case of $\mu > 0$ includes all rays that cover only increasing distance from the star, whereas $\mu < 0$ are rays that initially track paths that get closer to the star and only later extend to growing distance. Direct integration of equation~\eqref{eq:tauformula} then results in 
\begin{equation}
    \tau(\mu,r;\tau_*)=\tau_*\frac{R_*}{z_t}
    \begin{cases}
    t_-(\mu) & \mu\geq0\\
    t_+(|\mu|) + \pi & \mu<0
    \end{cases},\label{eq:opticaldepthfull}
\end{equation}
where
\begin{equation}
    t_\pm = \arctan{\left(\frac{R_*}{z_t}\right)} \pm \arctan{\left(\frac{\gamma}{\mu}\right)}.\label{eq:tau-t}
\end{equation}
We borrow the notation of $z_t$ from \citetalias{Cohen_X-ray_mass_loss_method} so that
\begin{equation}
    z_t = \sqrt{\left(1-\mu^2\right)r^2-R_*^2}\label{eq:zt}
\end{equation}
and define
\begin{equation}
    \gamma = \frac{R_*-r\left(1-\mu^2\right)}{z_t}.\label{eq:gammadef}
\end{equation}
See Appendix \mbox{\ref{sec:optical_depth_derivation}} for the full derivation of this expression. While not shown here, we verified that our optical depth is the same as used in \mbox{\citetalias{Cohen_X-ray_mass_loss_method}} -- \mbox{\citetalias{Cohen_zeta_pup_mass_loss_2020}}.

The novel approach to our model is to parametrize the locations where the heating occurs independently from the cool-wind emission-measure distribution, and to do this in such a way that fast gas can overtake slower gas right from the surface, and will continue to do so until it rams into something. This same approach was taken in \citet{Gayley_unified_hot-star_X-rays}. The idea that fast streams are initiated from surface perturbations and are only removed from the wind as they crash into slower gas causes the heating to exponentially fall off with radius, characterized by a constant probability per unit length
\begin{equation}
    \dv{P}{r}=\frac{1}{\ell_0},
\end{equation}
where $\ell_0$ is the second of our two parameters. A simple interpretation of $\ell_o$ is a mean-free path (in units of $R_*$) for a parcel of fast gas to overtake slow gas. Note this also implies a characteristic X-ray production radius $R_* + \ell_0$, which can be compared with $R_0$ parameter from \citetalias{Cohen_zeta_pup_mass_loss_2020} since both give a characteristic formation radius for any given line. In a mutually consistent picture, we might expect $R_*+\ell_0$ parameter to be somewhat larger than the \citetalias{Cohen_zeta_pup_mass_loss_2020} $R_0$ parameter, given that ours characterizes a ``turn-off" radius rather than a turn-on radius, but otherwise the quantities should be somewhat analogous.

Not all radii can be included in the integration of equation~\eqref{eq:luminositybase} due to the $\beta=1$ velocity law. Any given value of $\xi$ can only come from radii that are at least far enough from the star to get that much Doppler shift (if $\xi < 0$), and if $\xi > 0$, some of those rays would be occulted by the star. Because of this dependence on the sign of the $\xi$, we give the two cases separately below. From our definition of the angular dependent velocity coordinate, we can see that the minimum radius for $\xi<0$ is
\begin{equation}
    r_m(\xi\leq0)=\frac{R_*}{1+\xi}.
\end{equation}
The case of $\xi>0$ is more complicated because we have to include the effect of occultation. This involves solving for the minimum distance such that the X-ray is tangent to the star
\begin{equation}
    \mu = \mu_* = -\frac{\sqrt{r^2-R_*^2}}{r}.
\end{equation}
Substituting this into our $\xi$ coordinate, we then have to solve the quartic polynomial
\begin{equation}
    \left(\frac{r}{R_*}\right)^4 - w\left(\frac{r}{R_*}\right)^3+w\frac{r}{R_*} - \frac{1}{2}w=0,\label{eq:rminpoly}
\end{equation}
where $w=2/(1-\xi^2).$ This can be done using Ferrari's method, in which the real positive root takes the form
\begin{equation}
    r_m(\xi>0)=\left(
    \frac{w}{4}+S+\frac{1}{2} \sqrt{\frac{3}{4}w^2-4S^2-\frac{8w-w^3}{8S}}\right)R_*,\label{eq:rminback}
\end{equation}
where
\begin{align}
    S &= \frac{1}{2}\sqrt{\frac{1}{4}w^2+\frac{1}{3}\left(Q+\frac{\Delta_0}{Q}\right)},\\
    Q &= \left(\frac{1}{2}\Delta_1+\frac{1}{2}\sqrt{\Delta_1^2-4 \Delta_0^3}\right)^{1/3},\label{eq:Q}\\
    \Delta_0 &= 3w^2-6w,\\
    \Delta_1 &= 27\left(w^2-\frac{1}{2}w^3\right).\label{eq:Delta1}
\end{align}
See Appendix~\ref{sec:realcubicrootappendix} for details on the use of equation \eqref{eq:rminback}.

%%%%%%%%%%%%%%%%%%%%%%%%%%%%%%%%%%%%%%%%%%%%%%%%%%%%%%%%%%%%%%%%%%%%%%%%%%%%%%%%%%%%%%%%%%%%%%%%%%%%%%%%
%%%%%%%%%%%%%%%%%%%%%%%%%%%%%%%%%%%%%%%%%%%%%%%%%%%%%%%%%%%%%%%%%%%%%%%%%%%%%%%%%%%%%%%%%%%%%%%%%%%%%%%%

\section{Example Plots and Model Behavior}\label{sec:modelbehavior}

\begin{figure}
    \centering
    \includegraphics[width=\linewidth]{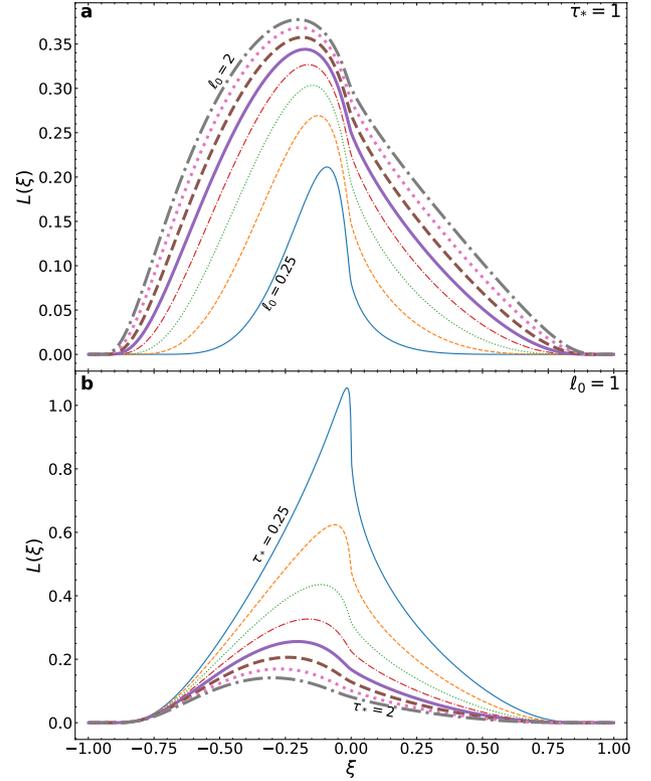}
    \caption{Model line profiles showing how equation~\eqref{eq:luminositybase} is affected by the parameters $\tau_*$ and $\ell_0$. (\textbf{a}) The parameter $\tau_* = 1$ is maintained as a constant while $\ell_0$ is varied. (\textbf{b}) The parameter $\ell_0 = 1$ is maintained as a constant while $\tau_*$ is varied. In both images, the varied parameter takes values ranging from 0.25 to 2 in steps of 0.25, and correspond to solid blue, dashed orange, dotted green, dotted-dashed red, thick purple, thick dashed brown, thick dotted pink, and thick dotted-dashed grey respectively.}
    \label{fig:example_profiles}
\end{figure}

We provide example plots of our model to show what effects our parameters have when isolated. In Fig.~\mbox{\ref{fig:example_profiles}}\textbf{a}, we fix $\tau_*=1$ and vary $\ell_0$ from 0.25 to 2 in steps of 0.25; the reverse situation is used for Fig.~\mbox{\ref{fig:example_profiles}}\textbf{b}. See the figure description for the legend of the plots. While most of the curves are very Gaussian-like (the scale of Fig.~\mbox{\ref{fig:example_profiles}}\textbf{b} belies the shape of the larger $\tau_*$ curves), for small $\tau_*$ the profile becomes very centrally peaked. This is an artefact of us allowing shocks to occur from the moment the gas leaves the surface. In the case where the overall wind can be parametrized as having a low optical depth, even photons from deep in the wind can easily escape.. However, this is not to say there is no significance to this. For example, if the solid blue curve of Fig.~\mbox{\ref{fig:example_profiles}}\textbf{b} were to be convolved with the \textit{Chandra} Line Spread Function, such a peak would be obscured due to the broadening and smoothing that would occur. It is thus possible that these low-velocity shocks are occurring, but either the photons do not escape from the low-lying radii or we do not have the resolution to distinguish them in over-all line feature. Further analysis will be needed to discover if these peaked profiles are reasonable deconvolutions of the observed line shapes given the ambiguity inherent in deconvolution methods.

Due to the complex nature of our model, we need to understand its response to small changes in parameters before attempting to fit it to real data. This will not only give us insight into how to analyze later results but also how to expect it to behave within spectral fitting software. To do this, we will look at the covariance between the parameters when specific observables are chosen as diagnostic quantities.

When parametric model fitting line profiles in X-ray data, it is common to use a normalized Gaussian profile, e.g.
\begin{equation}
    G(\lambda_0,\sigma,K) = \frac{K}{\sigma\sqrt{2\pi}}\exp\left(-\frac{1}{2}\left(\frac{\lambda-\lambda_0}{\sigma}\right)^2\right),
\end{equation}
such as in \citet{Pradhan_WR25_Gauss_Example}. Assuming the level of asymmetry present in a line profile is small enough that a Gaussian model can fit said profile statistically well, then the profile is sufficiently described by two quantities: the centroid $\lambda_0$ and width $\sigma$; modulo the total flux in the profile. These quantities are thus what we will use to test the covariance between our model parameters.

The centroid $C$ is the simpler of the two as it is simply the first moment of the profile distribution
\begin{equation}
    C \equiv \frac{\int_{-1}^1\xi L(\xi)\dd\xi}{\int_{-1}^1L(\xi)\dd\xi}.
\end{equation}
The width however, is not as simple. It is only true in the case of Gaussian distributions that the standard deviation $\sigma$ is also the width since the full width at half maximum is $\fwhm=2\sqrt{2\log(2))}\sigma$. This would not be case for our profiles, so we instead define the width $H$ to be the literal full width at half maximum
\begin{equation}
    H\equiv\xi_{R,\text{HM}}-\xi_{L,\text{HM}}.
\end{equation}
Here $\xi_{i,HM}$ are the right and left points of half-maximum for our model.

Discussing these observables should be done carefully with regard to our model since it does not produce perfectly Gaussian profiles. As is evident from Fig.~\ref{fig:example_profiles}, the model profile show both some skewness and tailedness (also commonly referred to as excess kurtosis). This can mean that the centroid and width, as defined above with regard to Gaussian models, may not necessarily be accurate nor descriptive enough of the model behavior. This does not mean there is no value in the discussion that follows. It is just one that should be considered a first order approximation to understand the model's global behavior.

\begin{figure*}
    \centering
    \includegraphics[width=\linewidth]{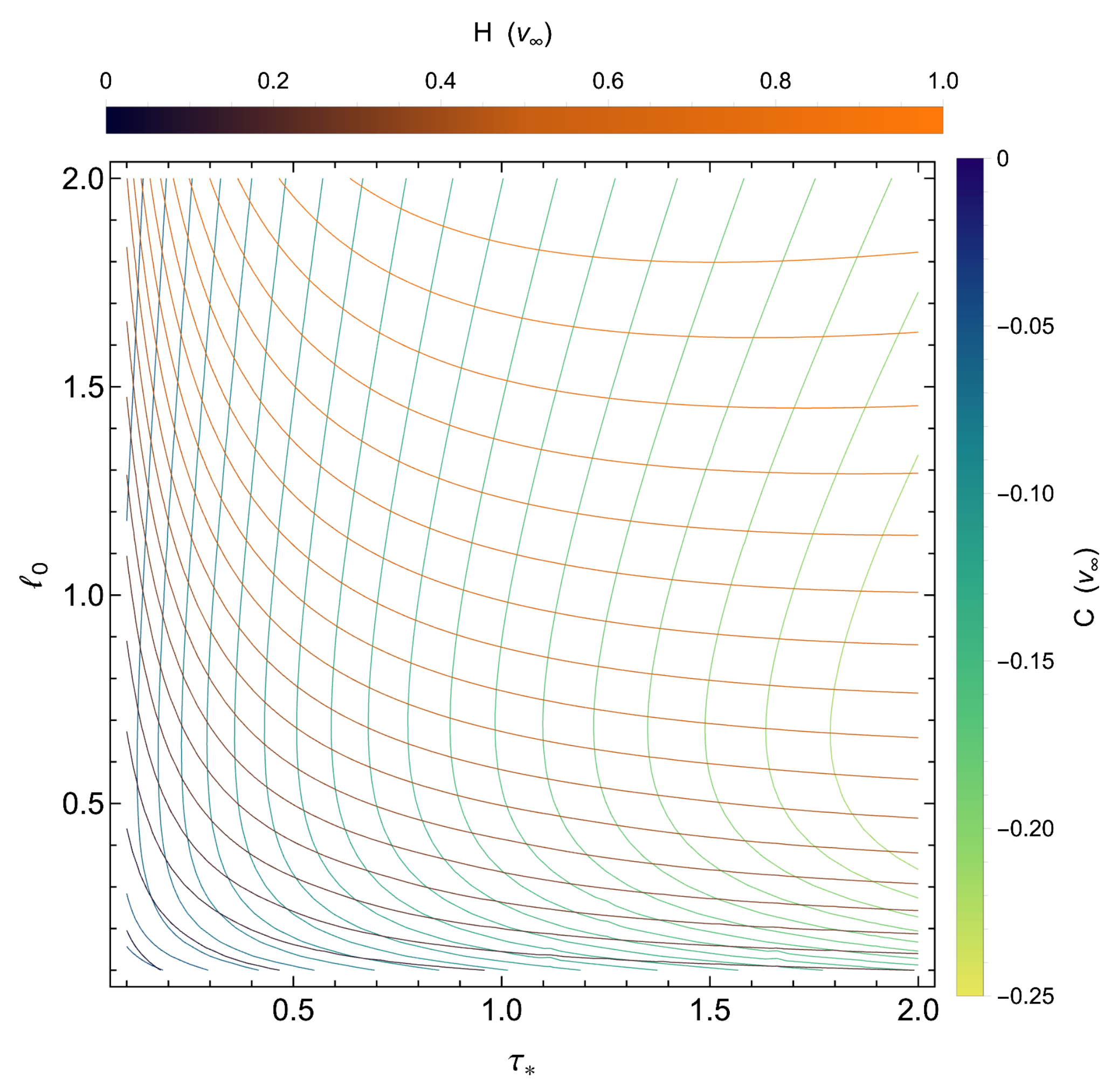}
    \caption{Contour plots of $C$ (the line shift measured relative to $v_\infty$) and $H$ (the line width measured relative to $v_\infty$) in the $(\tau_*, \ell_o)$ plane. The angle of intersection of the contours shows the covariance of the model parameters in each region. Regions where the contours meet at nearly right angles indicate parameter sets with good independence between the model parameters.}
    \label{fig:contourplot}
\end{figure*}

In a model of $N$ parameters and observable with no covariance, each observable would be effected by only one of the parameters and be independent of the others, i.e. \mbox{$O_i(a_1,...,a_N) = O_i(a_i)$}. Since we have two parameters and observables, we make the assertion that we can tie a line shape quantities behavior to a single parameter in this way. So to test our model's covariance, we plotted contours of $C$ and $H$ in Fig~\mbox{\ref{fig:contourplot}}. If our model has no covariance, then we expect the contours to be orthogonal; which we can see happens for one of two regions in the figure:

\begin{enumerate}
    \item $(\tau_*,\ell_0)\gtrsim(0.5,0.5)$: For this large region of parameter space, the contours form a near perfect grid, showing that to get large changes in $C$ you have to move in $\tau_*$ whereas large changes in $H$ are caused by motion in $\ell_0$.
    \item $\tau_*\lesssim0.5$ or $\ell_0\lesssim0.5$: When the parameters get small, though, there is ambiguity. In these regions, we can change either parameter by a small amount and notice a change in both observables.
\end{enumerate}

From this information, we can be confident that in large solution values there is minimal uncertainty between the parameters. However, if either parameter becomes small, then we can not be certain that our particular solution is unique nor well constrained. Additionally, we can assert the scale of the slopes for these regions. In the first region, we would expect a spectra fitting routine to have a steep slopes in both parameter directions. This would not be the case for small values though, as it may be shallow in one or both directions.

%%%%%%%%%%%%%%%%%%%%%%%%%%%%%%%%%%%%%%%%%%%%%%%%%%%%%%%%%%%%%%%%%%%%%%%%%%%%%%%%%%%%%%%%%%%%%%%%%%%%%%%%
%%%%%%%%%%%%%%%%%%%%%%%%%%%%%%%%%%%%%%%%%%%%%%%%%%%%%%%%%%%%%%%%%%%%%%%%%%%%%%%%%%%%%%%%%%%%%%%%%%%%%%%%

\section{Data and Modelling}\label{sec:data}

The observations for $\zeta$ Pup included all archival HETG data from 2018-2019, i.e. \textit{Chandra} Proposal Cycle 19, spanning a total exposure time of 813 ks. The list of observations with their Observation ID (Obs ID) are given in table~\ref{tab:ObsLog}. Each observation used the \textit{Chandra} HETGS, giving data sets associated with both of the two grating arrays: the medium energy grating (MEG) and high energy grating (HEG) \citep{HETGS_citation}.

\begin{table}
\centering
\caption{\textit{Chandra} Cycle 19 Observations of $\zeta$ Pup}\label{tab:ObsLog}
\begin{tabular}{lcc}
    \hline
    Obs ID & Exposure Time (ks) & Date\\
    \hline
    21113 & 17.72 & 2018 Jul 1\\
    21112 & 29.70 & 2018 Jul 2\\
    20156 & 15.51 & 2018 Jul 3\\
    21114 & 19.69 & 2018 Jul 5\\
    21111 & 26.86 & 2018 Jul 6\\
    21115 & 18.09 & 2018 Jul 7\\
    21116 & 43.39 & 2018 Jul 8\\
    20158 & 18.41 & 2018 Jul 30\\
    21661 & 96.88 & 2018 Aug 3\\
    20157 & 76.43 & 2018 Aug 8\\
    21659 & 86.35 & 2018 Aug 22\\
    21673 & 14.95 & 2018 Aug 24\\
    20154 & 46.97 & 2019 Jan 25\\
    22049 & 27.69 & 2019 Feb 1\\
    20155 & 19.69 & 2019 Jul 15\\
    22278 & 30.51 & 2019 Jul 16\\
    22279 & 26.05 & 2019 Jul 17\\
    22280 & 25.53 & 2019 Jul 20\\
    22281 & 41.74 & 2019 Jul 21\\
    22076 & 75.12 & 2019 Aug 1\\
    21898 & 55.70 & 2019 Aug 17\\
    \hline
\end{tabular}
\end{table} 

Each observation was retrieved from the \textit{Chandra} archive and reprocessed using the standard \textit{chandra\_repro} pipeline in \textsc{ciao} version 4.13. This process produced two first-order spectra for both the HEG and MEG arrays, corresponding to the positive and negative diffraction orders. For this analysis, we have chosen to focus solely on the MEG data as we are interested in the global shape of the lines. Each of the MEG observations were subsequently co-added to produce a single first order data set for each Obs ID and then summed to produce one data set using the \textit{combine\_grating\_spectra} command in \textsc{ciao}. This final data set was then rebinned by a constant factor of 3 before model fitting was done.

Using \textsc{xspec} version 12.12 \citep{Arnaud_xspec}, the emission lines were fit in narrow wavelength regions containing groups of overlapping or close features using our VBC model plus a continuum, all folded through the instrument response of the combined observation. We have specifically implemented our VBC model as a local Python (version 3.8) model in PyXspec, the Python interface for \textsc{xspec}. For the continuum, we used a constant model whose normalization was allowed to scale freely. For the emission lines, we have chosen to focus our model on a majority of the same lines that \citetalias{Cohen_zeta_pup_mass_loss_2020} fit. For lines that were too blended to be constrained independently, we restricted parameters by tying their values to a stronger line's; line flux was always a free parameter although they are not of interest for this analysis. For all the fits, we did not correct for absorption therefore the flux is as seen by the observer on Earth.  We assumed Poisson statistics and used a Cash (maximum likelihood) statistic, and computed confidence limits on the free parameters of interest.

Finally, to convert the wavelength scale of the detector to our Doppler shift coordinate, we used the conversion formula
\begin{equation}
    \xi = \left(\frac{\lambda}{\lambda_\mathrm{p}}-1\right)\frac{c}{v_{\infty,f}}.
\end{equation}
Here $\lambda_\mathrm{p}$ is the line's predicted rest wavelength as reported in AtomDB \citep{Smith_AtomDB,Foster_AtomDB}. We assume that the terminal velocity of the faster portions of $\zeta$ Pup's wind have a slightly higher terminal velocity than the unshocked portion of the wind. Specifically, we assert a conservative increase to $v_{\infty,f} = 2500 \text{ km s}^{-1}$ for the fast gas. We make this assumption on the grounds that in our VBC framework, the faster portion of the wind will have a steeper acceleration curve that would result in a faster terminal velocity than the slower and post-shock cooling portions of the wind seen in UV observations \citep{Puls_terminal_velocity}. Additionally, while a terminal velocity of $v_{\infty} = 2250 \text{ km s}^{-1}$ is the canonically accepted value for $\zeta$ Pup, \citet{Puls_terminal_velocity} does note that there can be up-to 10\% uncertainty in this value. Thus, our adoption of a slightly faster terminal speed for the fast portion is within the realm of possibility for even the slow gas.

\section{Results}\label{sec:results}

The model fitting results are summarized in Table~\ref{tab:modelresults} where we provide 68\% confidence intervals on our free parameters. Plots of these best fits are presented in Appendix~\ref{sec:all_model_fits} in the same order as they appear in the table. We can see from these plots that our assertion that the \textit{Chandra} response would smooth our model to be more Gaussian-like was accurate. In general though, asymmetries are still present and well modelled for all lines fit. Below we comment on fits to specific line features.

\begin{table}
\centering
\caption{Model Fitting Results \label{tab:modelresults}}
\begin{tabular}{lccc}
    \hline
    Line & $\lambda_\mathrm{p}$ (\AA) & $\tau_*$ & $\ell_0$ ($R_*$)\\
    \hline
    \ion{Si}{XIII} & 5.681 & $0.21_{-0.13}^{+0.15}$ & $0.90_{-0.19}^{+0.62}$\\
    \ion{Si}{XIV} & 6.180 & $0.07_{-0.04}^{+0.07}$ & $1.13_{-0.21}^{+0.23}$\\
    \ion{Si}{XIII} & 6.648, 6.688, 6.740 & $0.26_{-0.02}^{+0.02}$ & $1.71_{-0.05}^{+0.05}$\\
    \ion{Mg}{XII} & 8.419 & $0.38_{-0.06}^{+0.06}$ & $1.28_{-0.11}^{+0.06}$\\
    \ion{Mg}{XI} & 9.169, 9.231, 9.314 & $0.50_{-0.03}^{+0.03}$ & $1.18_{-0.03}^{+0.03}$\\
    \ion{Ne}{X} & 10.238 & $1.32_{-0.08}^{+0.19}$ & $0.81_{-0.07}^{+0.16}$\\
    \ion{Ne}{IX} & 11.544 & $0.94_{-0.15}^{+0.18}$ & $1.50_{-0.18}^{+0.21}$\\
    \ion{Ne}{X} & 12.132 & $1.37_{-0.09}^{+0.10}$ & $1.52_{-0.08}^{+0.09}$\\
    \ion{Fe}{XVII} & 15.014, 15.176, 15.261 & $1.0_{-0.18}^{+0.76}$ & $1.71_{-0.21}^{+0.60}$\\
    \ion{Fe}{XVII} & 16.78, 17.051, 17.096 & $2.75_{-0.36}^{+2.35}$ & $1.71_{-0.53}^{+0.89}$\\
    \hline
\end{tabular}
\end{table} 

\subsection{Comments on Specific Lines}
\subsubsection{He-like Lyman $\alpha$ Lines}

When fitting the \ion{Si}{XIII} and \ion{Mg}{XI} He Ly $\alpha$ triple lines, we modelled all three fir lines and tied the $\tau_*$ and $\ell_0$ values to the resonance line. While this ensured a physically realistic scenario for the emission and absorption of the X-rays, it was a challenge for \textsc{xspec}'s simplex algorithm to find a solution beyond the starting conditions. This was more of a problem for the \ion{Mg}{XI} line than \ion{Si}{XIII}, in which the normalization parameter appeared to be the more key parameter for finding a fit. More complex simplex algorithms, such as amoeba simplex in \textsc{isis}, or a detailed Markov Chain Monte Carlo analysis may find a fit that deviates from the starting parameters more. But even so, the residuals in Figure~\ref{fig:AllModelFits} and confidence intervals show that this starting position is close to the actual best-fit.

\subsubsection{\ion{Fe}{XVII} at 15 \AA}

Unlike \citetalias{Cohen_zeta_pup_mass_loss_2020}, we have chosen to model the entire line feature present at 15 -- 15.3 \AA. This feature consists of the 3C and 3D \ion{Fe}{XVII} emission lines at 15.014 and 15.261 \AA\ and the \ion{O}{VIII} H Ly $\gamma$ line at 15.176 \AA. As with the He-like triple lines, we tied the parameter values of these lines together. This feature showed the same difficult behavior with the simplex algorithm in moving away from the starting position. Here this may be due to the shallowness of the parameter space as the upper limits of the confidence intervals are quite large.

\subsubsection{\ion{Fe}{XVII} at 17 \AA}

This line feature is the last one that required the use of multiple model components as we model the 3F, 3G, and M2 \ion{Fe}{XVII} emission lines simultaneously. Again, these have their parameter values tied together, but unlike the previous ones there was no issue with moving off the starting value. The error associated with the best-fit values found are also significantly larger than the other lines, but this may be due to additional factors than the fiducial absorption considered here. As discussed in \citet{Grell_Fe17_ratio} (and with the authors in private correspondence), the flux ratio (M2+3G)/3F is anomalously low for $\zeta$ Pup as compared to other O stars and is potentially due to resonant Auger destruction of the 3G and/or M2 lines. This effect would then present itself as an enhancement in optical depth of these lines, potentially explaining the large upper bound on the confidence interval.

%%%%%%%%%%%%%%%%%%%%%%%%%%%%%%%%%%%%%%%%%%%%%%%%%%%%%%%%%%%%%%%%%%%%%%%%%%%%%%%%%%%%%%%%%%%%%%%%%%%%%%%%
%%%%%%%%%%%%%%%%%%%%%%%%%%%%%%%%%%%%%%%%%%%%%%%%%%%%%%%%%%%%%%%%%%%%%%%%%%%%%%%%%%%%%%%%%%%%%%%%%%%%%%%%

\section{Discussion}\label{sec:discuss}
\subsection{Spatial Heating Distribution}

\begin{figure}
    \centering
    \includegraphics[width=\linewidth]{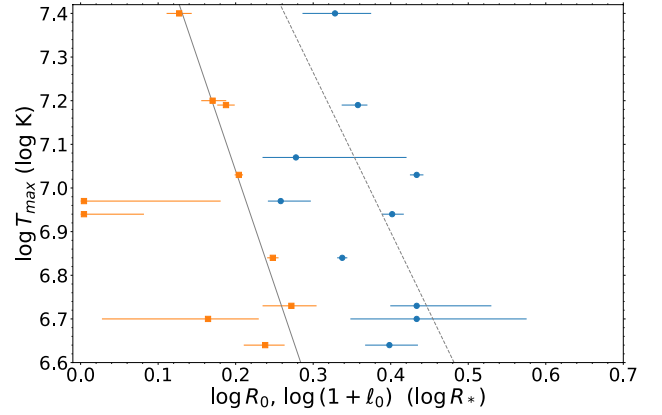}
    \caption{Comparison of $\ell_0$ from our model (blue dots) and $R_0$ from the \citetalias{Cohen_zeta_pup_mass_loss_2020} (orange squares). Both data sets are plotted against the temperature of maximum emissivity $T_{max}$ for their corresponding line as reported in AtomDB \citep{Smith_AtomDB,Foster_AtomDB}.}
    \label{fig:r0comparedR0}
\end{figure}

To understand the results of our fits, we will compare them against the corresponding values from \citetalias{Cohen_zeta_pup_mass_loss_2020}. Starting with the parameters $\ell_0$ in our model and $R_0$ from \citetalias{Cohen_zeta_pup_mass_loss_2020}, we've plotted in Fig.~\ref{fig:r0comparedR0} their logarithm vs the logarithm of the corresponding line's temperature of maximum emissivity $T_{max}$ reported in AtomDB \citep{Smith_AtomDB,Foster_AtomDB}. We use $T_{max}$ as a proxy for the post shock temperature, $kT_{shock}=\frac{3}{16}\mu m_\mathrm{H}v^2$, since we can expect a majority of a line's emission to be from plasma at or near that temperature. Because of this, $T_{max}$ is our parametrization of the shock strength, and its trends with $\ell_0$ and $R_0$ will tell us how the shock strength radially varies.

To our surprise, \textit{both} $\ell_0$ and $R_0$ show an inverse proportionality with the maximum temperature. To better illustrate this apparent trend, we've also plotted regression lines in Figure~\ref{fig:r0comparedR0}. If the LDI is the dominate mechanism, we would expect the $R_0$ trend to have a positive slope as the velocity evolution in \citet{Feldmeier_LDI_Non-linear_Growth} and \citet{Driessen_LDI_Sim} predict. This shows the robustness of our VBC theory for the fact that one of its main predictions, i.e., stronger shocks occurring closer to the stellar surface than weaker ones, is shown even when it is not invoked. This result is significant; especially in comparison to the results of \citetalias{Cohen_X-ray_mass_loss_method}, which found a relatively constant $R_0$ for all lines.

There are are two possible explanations for the changes in the distribution of $R_0$. As discussed in \citetalias{Cohen_zeta_pup_mass_loss_2020}, the X-ray emission from $\zeta$ Pup may have changed in 20 years; their evidence being their proposed 40\% increase in $\zeta$ Pup's mass-loss rate. Alternatively, the Cycle 1 observation may not have been long enough to get the signal-to-noise levels to an adequate level for the analysis. In which case, the same trend shown in Fig.~\ref{fig:r0comparedR0} would have appeared if the 813 ks were done during Cycle 1. There is difficulty in distinguishing between these two effects due to the changes in the instruments sensitivity. These changes have especially affected the longer wavelengths that would show the most change in optical depth due to increased material in the wind. Regardless of the reason for the trend differences between the two observations, the fact that the \citetalias{Cohen_zeta_pup_mass_loss_2020} results show an inverse proportionality is significant.

\subsection{Optical depth}

\begin{figure}
    \centering
    \includegraphics[width=\linewidth]{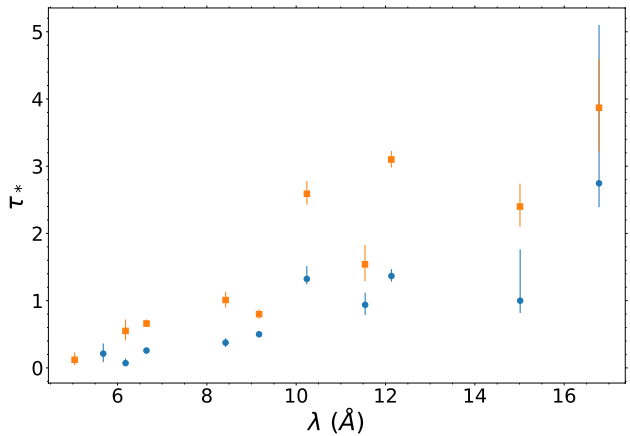}
    \caption{Comparison of the $\tau_*$ parameter from our model (blue dots) and those found by \mbox{\citetalias{Cohen_zeta_pup_mass_loss_2020}} (orange squares).}
    \label{fig:taustarcompare}
\end{figure}

For the optical depths $\tau_*$ from the two models, we've plotted them against the wavelengths of the associated lines in Fig.~\ref{fig:taustarcompare}. Again, our values are the blue dots and the orange squares are from \citetalias{Cohen_zeta_pup_mass_loss_2020}. We find that our values are on average 50\% smaller, implying a smaller mass-loss rate than was found in \citetalias{Cohen_zeta_pup_mass_loss_2020}. This result is both significant and troubling as it means that X-rays can \textit{not} be used as an independent tool for mass-loss determinations of hot stars. It appears that any mass-loss rates derived from X-ray line fitting are dependent on the specific heating model that is invoked in the derivation of the chosen line model. We make this case on the grounds that we have used a physically justifiable line model that uses a different heating model, yet we are able to fit the X-ray lines just as well.

Of all the assumptions used in our model, our choice of fast gas terminal speed could be the cause of the lower $\tau_*$ values we found. While we did not explore the effects of having the fast gas terminal speed $v_{\infty,f}$ be a free parameter, we did find that lowering it to the canonical $2250 \text{ km s}^{-1}$ caused an of order 10 -- 20\% increase to $\tau_*$. However, even this would not be enough to explain the discrepancy between our derived optical depths and those found in \citetalias{Cohen_zeta_pup_mass_loss_2020}. Additionally, lowering $v_{\infty,f}$ would then require that the prevailing cold gas terminal speed $v_\infty$ be decreased as well. As a result, we would infer an increase to \citetalias{Cohen_zeta_pup_mass_loss_2020}'s $\tau_*$ values as well since $\tau_* \propto v_\infty^{-1}$. For this reason we argue that the lower optical depth values we find are real results of our heating parametrization, not artefacts of our model assumptions.

\subsection{Future Work}
Because both models use a low-order parameterization, there is considerable ambiguity in which line profile model is the better reflection of reality. We thus propose a test of their reliability using the upcoming \textit{XRISM} observatory. If $\zeta$ Pup or any other O star is a future target for that observatory, that observational data could be used as laboratory for testing these two models. If the models' derived mass-loss rates differ at a statistically significant level, then the heating-model dependence is unavoidable. And if the spatial heating distributions also show an inverse correlation between radius and formation temperature, that provides evidence that the LDI alone cannot explain the shocking of the gas, without priming via boundary variations in the mass flux.

We now provide a short list of issues that should be addressed in future work.

\begin{enumerate}
    \item While \textit{XRISM} will provide a stronger test, we can still use the current \textit{Chandra} and \textit{XMM-Newton} archives to test our VBC model. Using the same ensemble of stars from \citet{Cohen_mass_loss_archive}, this analysis can test if the discrepancies between the derived $\tau_*$ values is true for all O stars or only $\zeta$ Pup.
    
    \item Fitting our model to $\zeta$ Pup's Cycle 1 data could provide a test for changes in its X-ray emission in two ways. First, it may be able to confirm the increase in the $\tau_*$ values as any real mass-loss rate change will be independent of the model choice. Second, if we were to find that our spatial parameter mirrors the \mbox{\citetalias{Cohen_X-ray_mass_loss_method}} results as it does for \mbox{\citetalias{Cohen_zeta_pup_mass_loss_2020}}'s, then that would confirm that the X-ray production has changed. However, if we still find an inverse proportionality between the maximum temperature and radial parameter, then then it would be the case that the Cycle 1 data has too low of signal-to-noise for \textit{windprof} and \textit{hewind} to give meaningful results.

    \item Shocked high-velocity gas should move faster than the prevailing wind clumps it runs into, so the wind terminal velocity derived from UV analysis \citep{Puls_terminal_velocity} should underestimate the maximum fast-gas speed. Allowing for the terminal velocity of the fast gas $v_{\infty,f}$ to be a free parameter is a needed avenue of exploration for future work with our model. Both to constrain the maximum speed and also to understand how it affects the model rigorously.
    
    \item  Our assumption of spherical X-ray emission omits detailed structures responsible for 1.78 and 2.56 day periodicities seen in $\zeta$ Pup's photometric and emission spectra as discussed in \citet{Howarth_zeta_Pup_props} and \citet{Nichols_zeta_Pup_period}. These features speak to coherent structures in the wind that can be analyzed as an additional shock mechanism, whose overall properties could be similar or quite different from the ones we fit.  Future work should account for these structures by including corrections to the absorbing optical depth or the heating distribution.
    
    \item  Our assumption of isotropic emission goes beyond the assumption of a spherical wind, it also requires that the X-ray lines be optically thin.  Future models should consider the potential effects of optically thick resonance scattering.
    
\end{enumerate}

%%%%%%%%%%%%%%%%%%%%%%%%%%%%%%%%%%%%%%%%%%%%%%%%%%%%%%%%%%%%%%%%%%%%%%%%%%%%%%%%%%%%%%%%%%%%%%%%%%%%%%%%
%%%%%%%%%%%%%%%%%%%%%%%%%%%%%%%%%%%%%%%%%%%%%%%%%%%%%%%%%%%%%%%%%%%%%%%%%%%%%%%%%%%%%%%%%%%%%%%%%%%%%%%%

\section{Conclusions}\label{sec:conclusions}

We have fit a new X-ray line model to recent 813 ks \textit{Chandra} X-ray line data on $\zeta$ Puppis to offer an alternative parametrization of the spatial heating distribution within the winds of O-type stars. This alternative distribution is inspired by variable boundary conditions and efficient radiative cooling, rather than shocks internally generated by the LDI that cool slowly enough to retain the overall emission measure structure of a spherically expanding wind that remains hot, as used in \citetalias{Cohen_X-ray_mass_loss_method} -- \citetalias{Cohen_zeta_pup_mass_loss_2020}. We find that our model is also able to fit the observed line shapes, but leads to an alternative estimate of the wind mass-loss rate that is lower by about a factor of 2 (cf.  Fig.~\ref{fig:taustarcompare}).

We conclude from this that using X-ray reabsorption to infer a clumping-independent mass-loss rate is not robust to the assumed details of the heating structure, forcing the user to choose theoretical preferences about the nature of the heating. If one subscribes to the picture of a “turn-on radius” for X-ray heating, the method of \citetalias{Cohen_X-ray_mass_loss_method} may be used and larger mass-loss rates inferred, whereas if one prefers a picture where the shock heating can begin almost immediately above the photosphere, tapering off with radius as the initially accelerated gas overtakes slower gas and is removed from the X-ray generating population by efficient cooling, then this new approach can also succeed in matching the data. Hence more attention is needed to constrain the nature of the heating in order for X-ray reabsorption to be used as a mass-loss rate diagnostic.

Each of these approaches to hot-gas parametrization come from different causes of the strong shocks, and the consistency of the inferred spatial distribution with these different causes can be used to inform preferences between the models. In particular, we find that the characteristic radius at which each X-ray temperature appears is roughly inversely proportional to that temperature. This trend holds for both our heating parametrization and for the one used in \citetalias{Cohen_zeta_pup_mass_loss_2020}, as shown in Fig.~\ref{fig:r0comparedR0}. This trend is seen more clearly in the expanded data set involving the most recent \textit{Chandra} observations, which may be why it was not originally recognized in \citetalias{Cohen_X-ray_mass_loss_method}. It should be noted that this trend cannot be explained by the simple fact that harder X-rays escape from deeper in the wind, because our profile fitting includes this fact yet still requires that harder lines form generally closer to the star. This trend is not expected if these strong shocks are due to the LDI, but does seem more natural if large shock jumps are associated with rapid acceleration of wind streams that were initially under-loaded right from the surface. To strengthen these conclusions, heating parametrizations for the purposes of profile fitting must be coupled more closely to realistic X-ray generation mechanisms from hydrodynamic shock models.

These conclusions also relate to evidence that the mass-loss rate of $\zeta$ Pup may have quite recently increased by some 40\%. If the heating affects the inferred mass flux, then changes in the latter could be due to changes in the former. In addition, the discovered correlation between X-ray temperature and radius allows another way to search for secular changes in the wind of $\zeta$ Pup from \textit{Chandra} cycle to cycle, to look for changes in the nature of the heating.

%%%%%%%%%%%%%%%%%%%%%%%%%%%%%%%%%%%%%%%%%%%%%%%%%%%%%%%%%%%%%%%%%%%%%%%%%%%%%%%%%%%%%%%%%%%%%%%%%%%%%%%%
%%%%%%%%%%%%%%%%%%%%%%%%%%%%%%%%%%%%%%%%%%%%%%%%%%%%%%%%%%%%%%%%%%%%%%%%%%%%%%%%%%%%%%%%%%%%%%%%%%%%%%%%

\section*{Acknowledgements}
The scientific results in this article are based on data retrieved from the \textit{Chandra} Data Archive, software provided by the Chandra X-ray Center (CXC) in the application packages \textsc{ciao}, and software provided by NASA's High Energy Astrophysics Science Archive Research Center (HEASARC) in the application \textsc{xspec}. Support for NAM was provided by NASA Chandra General Observer Program Cycle 19 grant GO8-19011D, a Chandra Research Visitor Award, and the University of Wisconsin-Eau Claire Office of Research and Sponsored Programs through the URCA and sabbatical programs. Support for DPH was provided by NASA through the Smithsonian Astrophysical Observatory (SAO) contract SV3-73016 to MIT for Support of the Chandra X-Ray Center (CXC) and Science Instruments. CXC is operated by SAO for and on behalf of NASA under contract NAS8-03060. We thank our reviewer Achim Feldmeier for his comments and suggestions that significantly improved this paper.

%%%%%%%%%%%%%%%%%%%%%%%%%%%%%%%%%%%%%%%%%%%%%%%%%%%%%%%%%%%%%%%%%%%%%%%%%%%%%%%%%%%%%%%%%%%%%%%%%%%%%%%%
%%%%%%%%%%%%%%%%%%%%%%%%%%%%%%%%%%%%%%%%%%%%%%%%%%%%%%%%%%%%%%%%%%%%%%%%%%%%%%%%%%%%%%%%%%%%%%%%%%%%%%%%

\section*{Data Availability}
The X-ray spectral data used in this article are available in the \textit{Chandra} Data Archive at \url{https://cxc.cfa.harvard.edu/cda/}. Observations are uniquely identified by an observation identifier (Obs ID) given in Table~\ref{tab:ObsLog}.

%%%%%%%%%%%%%%%%%%%%%%%%%%%%%%%%%%%%%%%%%%%%%%%%%%%%%%%%%%%%%%%%%%%%%%%%%%%%%%%%%%%%%%%%%%%%%%%%%%%%%%%%
%%%%%%%%%%%%%%%%%%%%%%%%%%%%%%%%%%%%%%%%%%%%%%%%%%%%%%%%%%%%%%%%%%%%%%%%%%%%%%%%%%%%%%%%%%%%%%%%%%%%%%%%

\bibliographystyle{mnras}
\bibliography{BIB}

%%%%%%%%%%%%%%%%%%%%%%%%%%%%%%%%%%%%%%%%%%%%%%%%%%%%%%%%%%%%%%%%%%%%%%%%%%%%%%%%%%%%%%%%%%%%%%%%%%%%%%%%
%%%%%%%%%%%%%%%%%%%%%%%%%%%%%%%%%%%%%%%%%%%%%%%%%%%%%%%%%%%%%%%%%%%%%%%%%%%%%%%%%%%%%%%%%%%%%%%%%%%%%%%%

\appendix

\section{Optical Depth Derivation}\label{sec:optical_depth_derivation}

For a photon emitted at a radius $r$ from the star at an angle $\mu$, its radial coordinate is
$r' = \sqrt{r^2+s^2+2rs\mu}$. If we make a change of integration variable to $\dd r'$ in equation~\eqref{eq:tauformula}, then for positive direction cosine the optical depth is
 \begin{equation}
    \tau_+=\tau(\mu\geq0)=\tau_*\frac{R_*}{z_t}\left(\arctan{\left(\frac{R_*}{z_t}\right)}-\arctan{\left(\frac{\gamma}{\mu}\right)}\right),
 \end{equation}
where $z_t$ and $\gamma$ are defined by equations~\eqref{eq:zt} and \eqref{eq:gammadef}.

For the case of negative direction cosine,  the photon will still travel through the optical depth of the positive case, so only an extra amount $\Delta\tau$ needs to be added:
\begin{equation}
    \tau_-=\tau(\mu<0)=\tau_+(|\mu|)+\Delta\tau,
\end{equation}
where
\begin{equation}
    \Delta\tau=2\tau_*\int_0^{|\mu|r}\frac{R_*v_\infty}{r''^2v(r'')}\dd s,
\end{equation}
and $r''=\sqrt{r^2+s^2-2rs|\mu|}$ is the radial coordinate of the photon when behind the star. Following a similar procedure as before, we change the integration variable to $\dd r''$ and add the result to the positive case to get
\begin{equation}
    \tau_-=\tau_*\frac{R_*}{z_t}\left(\arctan{\left(\frac{R_*}{z_t}\right)}+\arctan{\left(\frac{\gamma}{|\mu|}\right)}+\pi\right).
\end{equation}

%%%%%%%%%%%%%%%%%%%%%%%%%%%%%%%%%%%%%%%%%%%%%%%%%%%%%%%%%%%%%%%%%%%%%%%%%%%%%%%%%%%%%%%%%%%%%%%%%%%%%%%%
%%%%%%%%%%%%%%%%%%%%%%%%%%%%%%%%%%%%%%%%%%%%%%%%%%%%%%%%%%%%%%%%%%%%%%%%%%%%%%%%%%%%%%%%%%%%%%%%%%%%%%%%

\section{Occultation Minimum Radius}\label{sec:realcubicrootappendix}

Special care must be taken when computing the roots of equation~\eqref{eq:rminpoly} as Ferrari's method allows for any of the three cubic roots given by equation~\eqref{eq:Q} to be used. The form of the positive real root equation~\eqref{eq:rminback} makes use of the principal cubic root, i.e., the cubic root with the largest real part. This form is then best used in software with access to complex arithmetic. We give a short explanation below for the form of $r_m(\xi>0)$ to use if complex arithmetic is undesired or not possible to use. 

Let
\begin{equation}
    Q_R = \sqrt[3]{\frac{1}{2}\Delta_1+\frac{1}{2}\sqrt{\Delta_1^2-4\Delta_0^3}},
\end{equation}
where we define the operation $\sqrt[3]{x}$ to always give the real cubic root. In this case, the four roots of the quartic polynomial are
\begin{equation}
    \begin{aligned}
        r_{1,2}=\left(\frac{w}{4}-S\pm\frac{1}{2} \sqrt{\frac{3}{4}w^2-4S^2+\frac{8w-w^3}{8S}}\right)R_*\\
        r_{3,4}=\left(\frac{w}{4}+S\pm\frac{1}{2} \sqrt{\frac{3}{4}w^2-4S^2-\frac{8w-w^3}{8S}}\right)R_*\\
    \end{aligned}
\end{equation}
If the realness of these functions is checked for the region of $0<\xi<1$, we find that they are piece-wise real in pairs of $(r_1,r_3)$ and $(r_2,r_4)$ with a discontinuity at $\xi=\sqrt{1-\frac{1}{\sqrt{2}}}$. Of these two pairs, only $(r_1,r_3)$ are positive. Thus when only real values are desired, the form of $r_m(\xi>0)$ to use is
\begin{equation}
    r_{m,R}(\xi>0)=
    \begin{cases}
    r_1, & \xi<\sqrt{1-\frac{1}{\sqrt{2}}}\\
    r_3, & \xi\geq\sqrt{1-\frac{1}{\sqrt{2}}}
    \end{cases}\label{rminreal}.
\end{equation}

%%%%%%%%%%%%%%%%%%%%%%%%%%%%%%%%%%%%%%%%%%%%%%%%%%%%%%%%%%%%%%%%%%%%%%%%%%%%%%%%%%%%%%%%%%%%%%%%%%%%%%%%
%%%%%%%%%%%%%%%%%%%%%%%%%%%%%%%%%%%%%%%%%%%%%%%%%%%%%%%%%%%%%%%%%%%%%%%%%%%%%%%%%%%%%%%%%%%%%%%%%%%%%%%%

\section{All Model Fits}\label{sec:all_model_fits}

\begin{figure*}
    \centering
    \includegraphics[width=0.8\linewidth]{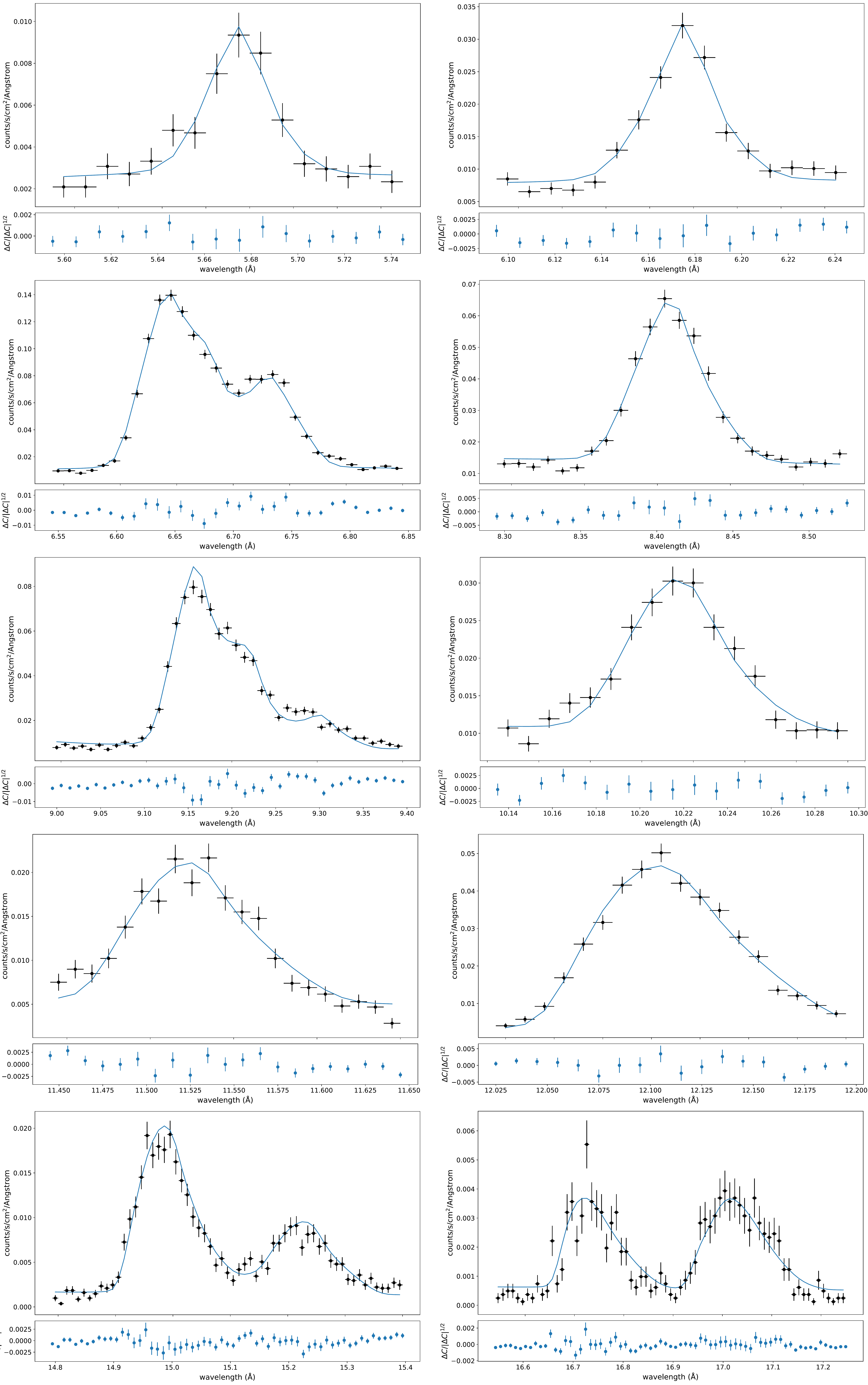}
    \caption{Best fit plots of the VBC model to the total summed MEG data given in the same order as Table~\ref{tab:modelresults}.}
    \label{fig:AllModelFits}
\end{figure*}

\pagebreak
%%%%%%%%%%%%%%%%%%%%%%%%%%%%%%%%%%%%%%%%%%%%%%%%%%%%%%%%%%%%%%%%%%%%%%%%%%%%%%%%%%%%%%%%%%%%%%%%%%%%%%%%
%%%%%%%%%%%%%%%%%%%%%%%%%%%%%%%%%%%%%%%%%%%%%%%%%%%%%%%%%%%%%%%%%%%%%%%%%%%%%%%%%%%%%%%%%%%%%%%%%%%%%%%%

\bsp	% typesetting comment
\label{lastpage}
\end{document}